\documentclass{PoS}

\usepackage[utf8]{inputenc}
\usepackage{amsmath,overpic}
\usepackage[href]{journals}

\title{Searching for New Physics through Charm at CDF}

\ShortTitle{Searching for New Physics through Charm at CDF}

\author{Angelo Di Canto%
	\thanks{Speaker on behalf of the CDF Collaboration.}
        \thanks{I would like to thank the BEAUTY 2011 organizers for the opportunity of speak at this conference
        and the colleagues of the CDF Collaboration for assisting me while preparing the talk and this document.}\\
        INFN \& University of Pisa, Fermilab\\
       E-mail: \email{angelo.dicanto@pi.infn.it}}

\abstract{We report a search for non-Standard Model physics through the measurement of CP-violating asymmetry in the Cabibbo suppressed \mbox{$D^0\to\pi^+\pi^-$} and \mbox{$D^0\to K^+K^-$} decays reconstructed in about 5.94 fb$^{-1}$ of CDF data. We use the strong $D^{\star +}\to D^0\pi^+$ decay  (``$D^{\star}$ tag'') to identify the flavor of the charmed meson at production time and exploit CP-conserving strong $c\bar{c}$ pair-production in $p\bar{p}$ collisions. Large samples of Cabibbo favored $D^0\to K^-\pi^+$ decays with and without $D^{\star}$ tag are used to highly suppress systematic uncertainties due to detector effects. The results are the world's most precise measurements to date.}

\FullConference{The 13th International Conference on B-Physics at Hadron Machines\\
		 April 4-8 2011\\
		 Amsterdam, The Netherlands}

\newcommand{\Acp}{\ensuremath{A_{\text{CP}}}}
\newcommand{\acp}[1]{\ensuremath{a_{\text{CP}}^{\text{#1}}}}
\newcommand{\Acpraw}{\ensuremath{\Acp^{\text{raw}}}}
\newcommand{\stat}{\ensuremath{\mathit{~(stat.)}}}
\newcommand{\syst}{\ensuremath{\mathit{~(syst.)}}}
\newcommand{\Dbar}{\ensuremath{\overline{D}{}}}

\begin{document}
\graphicspath{{./figs/}}

\section{Introduction}
One of the outstanding problems in particle physics is that the Standard Model (SM) implementation of CP violation, through the presence of the Cabibbo-Kobayashi-Maskawa phase, produces effects that are far from sufficient to  explain the matter-antimatter asymmetry of the Universe. While investigations of the $K$ and $B$ systems have and will continue to play a central role in our quest to understand flavor physics and CP violation, in-depth examinations of the $D$ mesons sector have yet to be performed with enough precision, leaving a gap in our knowledge. Since charm is the only heavy charged $+2/3$ quark presently accessible to experiment, it provides the sole window of opportunity to examine flavor physics in this sector that is complementary to the one of down-type quarks. Examples of clean channels with possible additional sources of CP violation in the charm system are the singly-Cabibbo suppressed transitions such as $D^0\to\pi^+\pi^-$ and $D^0\to K^+ K^-$. Contribution to these decays from ``penguin'' amplitudes are negligible in the SM, so the presence of New Physics (NP) particles could enhance the size of CP violation with respect to the SM expectation. Any asymmetry significantly larger than a few times $0.1\%$ may unambiguously indicate NP contributions \cite{theory}.

We present a high statistics search for CP violation in the $D^0\to\pi^+\pi^-$ and $D^0\to K^+ K^-$ decays, collectively referred to as $D^0\to h^+h^-$ in this paper, through the measurement of the time-integrated CP asymmetry:
\begin{equation}\label{eq:acp}
\Acp(h^+h^-)=\frac{\Gamma(D^0\to h^+h^-)-\Gamma(\Dbar^0\to h^+h^-)}{\Gamma(D^0\to h^+h^-)+\Gamma(\Dbar^0\to h^+h^-)}
\approx \acp{dir} + \frac{\langle t \rangle}{\tau}\acp{ind}.
\end{equation}
This asymmetry, owing to the slow mixing rate of charm mesons, is to first order the linear combination of a direct, $\acp{dir}$, and an indirect, $\acp{ind}$, term through a coefficient that is the mean proper decay time of $D^0$ candidates, $\langle t \rangle$, in units of $D^0$ lifetime ($\tau \approx 0.4$ ps). Since the value of $\langle t \rangle$ depends on the observed proper time distribution, different experiments may measure different values of $\Acp(h^+h^-)$. The measurement, described with further details in \cite{cdfnote}, has been performed on about $5.94$~fb$^{-1}$ of $p\bar{p}$ collisions at $\sqrt{s}=1.96$ TeV recorded by the CDF~II detector at Fermilab's Tevatron collider.

\section{Analysis overview}
We measure the asymmetry using $D^0\to h^+h^-$ decays from charged $D^\star$ mesons through fits of the $D^0\pi$ mass distributions. The observed asymmetry includes a possible contribution from actual CP violation, diluted in much larger effects from instrumental charge-asymmetries. We exploit a fully data-driven method that uses higher statistic samples of $D^\star$-tagged (indicated with an asterisk) and untagged Cabibbo-favored $D^0\to K^-\pi^+$ decays to correct for all detector effects, thus suppressing systematic uncertainties to below the statistical ones. The uncorrected ``raw'' asymmetries \footnote{``Raw'' are the observed asymmetries in signal yields, $$\Acpraw(D^0\to f) = \frac{N_{\text{obs}}(D^0\to f)-N_{\text{obs}}(\Dbar^0\to\bar{f})}{N_{\text{obs}}(D^0\to f)+N_{\text{obs}}(\Dbar^0\to\bar{f})},$$before any correction for instrumental effects has been applied.} in the three samples can be written as a sum of several contributions:
\begin{align*}
\Acpraw(hh^\star) &= \Acp(hh) + \delta(\pi_s)^{hh^\star},\\
\Acpraw(K\pi^\star) &= \Acp(K\pi) + \delta(\pi_s)^{K\pi^\star} + \delta(K\pi)^{K\pi^\star},\\
\Acpraw(K\pi) &= \Acp(K\pi) + \delta(K\pi)^{K\pi},
\end{align*}
where
\begin{itemize}
\item $\Acp(hh)$ and  $\Acp(K\pi)$ are the actual physical asymmetries; 
\item $\delta(\pi_s)^{hh^\star}$ and $\delta(\pi_s)^{K\pi^\star}$ are the instrumental asymmetries in reconstructing a positive or negative soft pion associated to a $h^+h^-$ and a $K^\mp\pi^\pm$ charm decay. This is mainly induced by charge-asymmetric track-reconstruction efficiency/absorption rates at low transverse momentum.
\item $\delta(K\pi)^{K\pi}$ and $\delta(K\pi)^{K\pi^\star}$ are the instrumental asymmetries in reconstructing a $K^\mp\pi^\pm$ charm decay for the untagged and the $D^\star$-tagged case respectively. These are mainly due to the difference in interaction cross-section with matter between positive and negative kaons. Smaller effects are due to charge-curvature asymmetries in track triggering and reconstruction.
\end{itemize}
The physical asymmetry is extracted by subtracting the instrumental effects through the combination
\begin{equation}\label{eq:formula}
\Acp(hh) = \Acpraw(hh^\star) - \Acpraw(K\pi^\star) + \Acpraw(K\pi),
\end{equation}
Any instrumental effect can vary as a function of a number of kinematic variables or environmental conditions in the detector, but if the kinematic distributions of soft pions are consistent in $K\pi^\star$ and $hh^\star$ samples, and the distributions of $D^0$ decay products are consistent in $K\pi^\star$ and $K\pi$ samples, then $\delta(\pi_s)^{hh^\star} \approx \delta(\pi_s)^{K\pi^\star}$ and $\delta(K\pi)^{K\pi^\star}\approx \delta(K\pi)^{K\pi}$ and the above relation is valid. This condition was verified in the analysis by inspecting a large set of kinematic distributions and applying small corrections (reweight) when needed.

\section{Measurement}
The trigger selects a pair of tracks from oppositely charged particles that have a distance of closest approch to the beamline (impact parameter) inconsistent with having originated from the primary vertex. We reconstruct signals consistent with the desired two-body decays ($h^+h^-$ or $K^-\pi^+$ or $K^+\pi^-$) of a neutral charmed meson ($D^0$ or $\Dbar^0$). To remove most non-promptly produced charmed mesons we also require the impact parameter of the $D^0$ candidate not to exceed $100\ \mu$m. Then we associate a low-momentum charged particle to the meson candidate to construct a $D^{\star+}$ (or $D^{\star-}$) candidate. The flavor of the charmed meson is determined from the charge of the pion in the strong \mbox{$D^{\star+}\to D^0\pi^+$} (or \mbox{$D^{\star-}\to \Dbar^0\pi^-$}) decay. Sample-specific mass requirements are used for the two tagged samples: we ask the two-body invariant mass to lie within $24$ MeV/c$^2$ of the nominal $D^0$ mass, corresponding to about $3\sigma$ around the peak. We reconstruct approximately $215^\cdot000$ $D^\star$-tagged $D^0\to\pi^+\pi^-$ (and charge conjugated) decays, $476^\cdot000$ $D^\star$-tagged $D^0\to K^+K^-$ (and c. c.) decays, 5 million $D^\star$-tagged $D^0\to\pi^+K^-$ (and c. c.) decays and 29 million $D^0\to\pi^+K^-$ (and c. c.) decays where no tag was required.

\begin{figure}[p]
\centering
\begin{overpic}[height=0.24\textheight,grid=false,trim=0 160pt 0 0,clip]{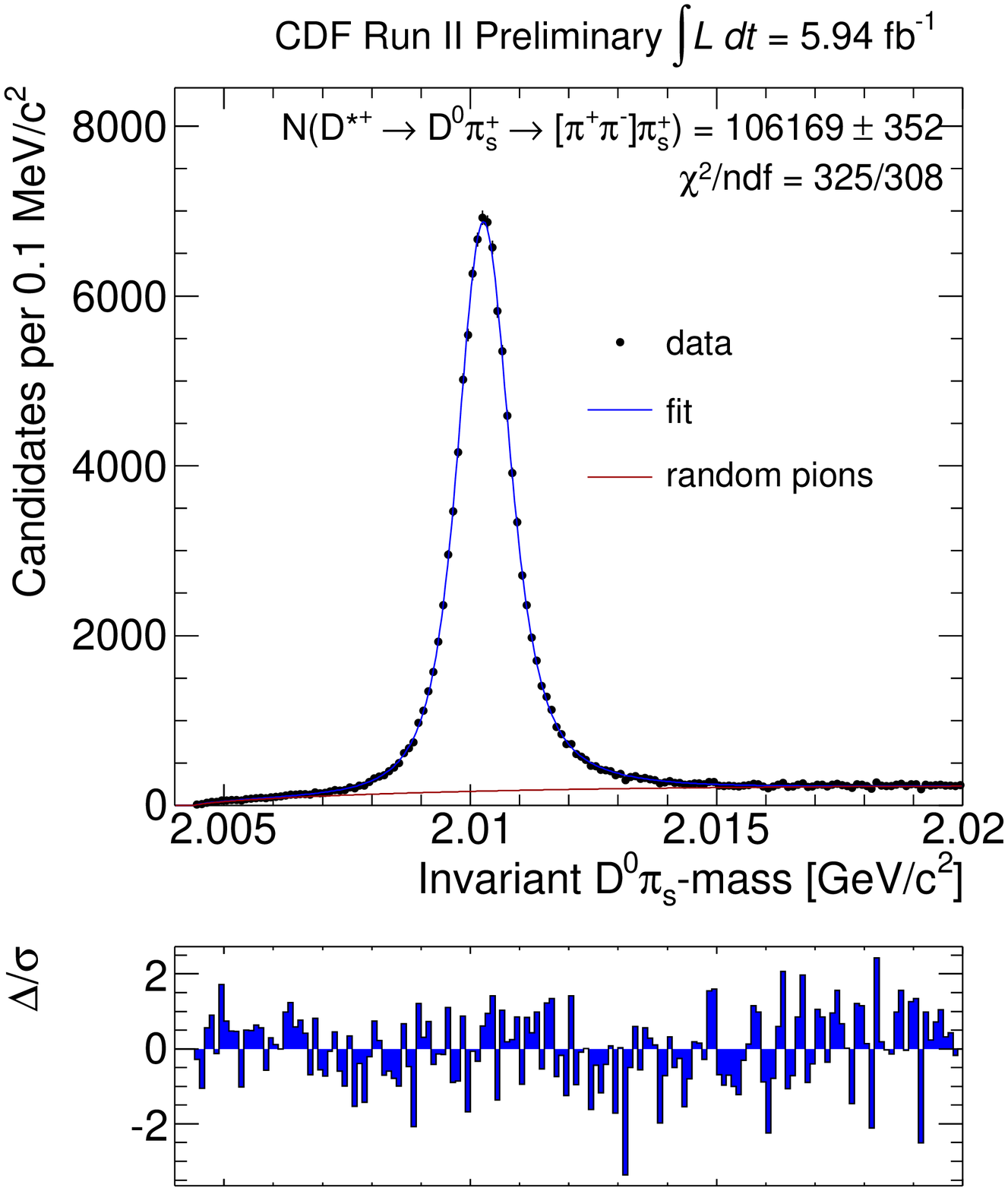}
\put(82,22){(a)}
\end{overpic}\hfil
\begin{overpic}[height=0.24\textheight,grid=false,trim=0 160pt 0 0,clip]{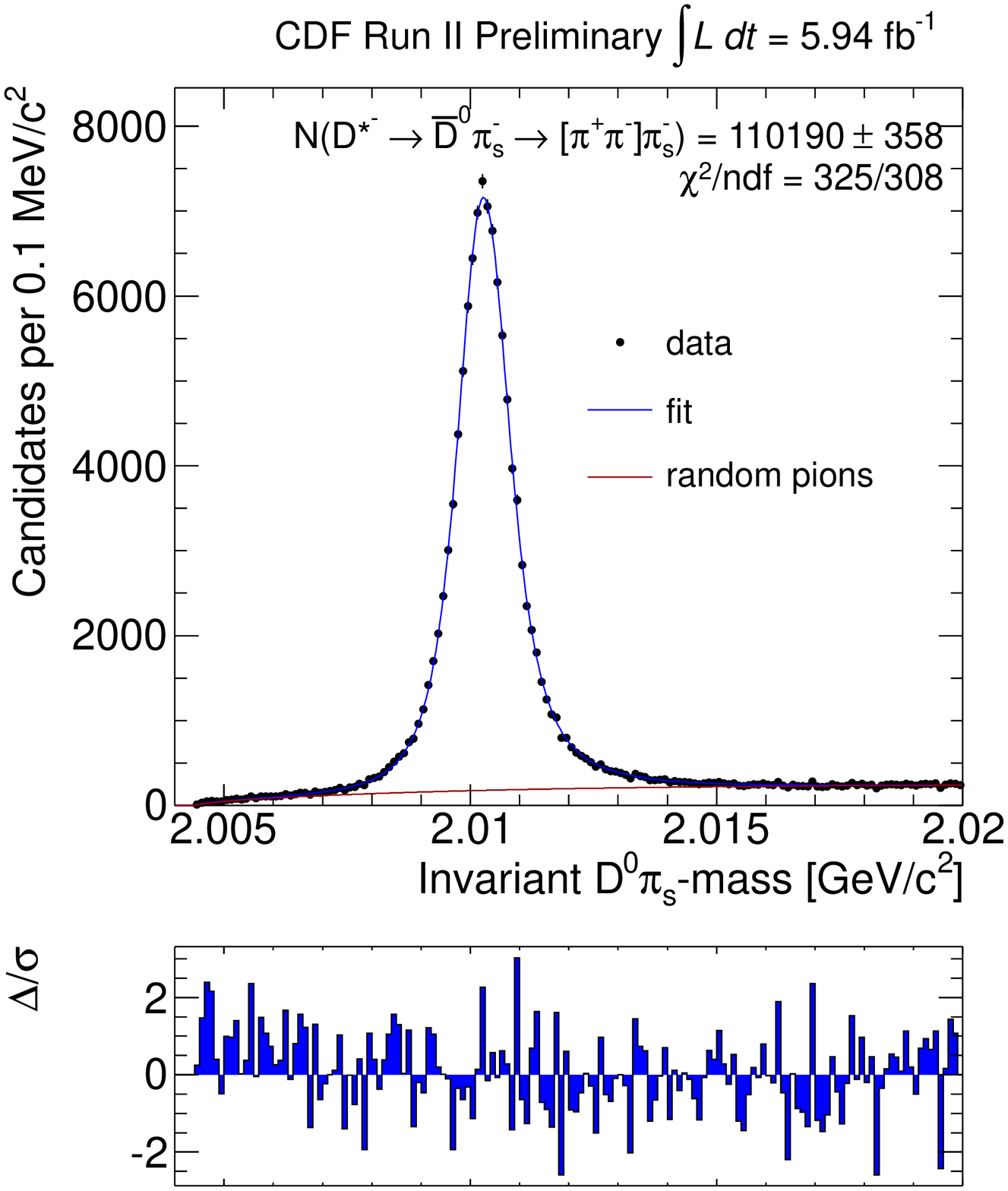}
\put(82,22){(b)}
\end{overpic}\vfill
\begin{overpic}[height=0.24\textheight,grid=false,trim=0 160pt 0 0,clip]{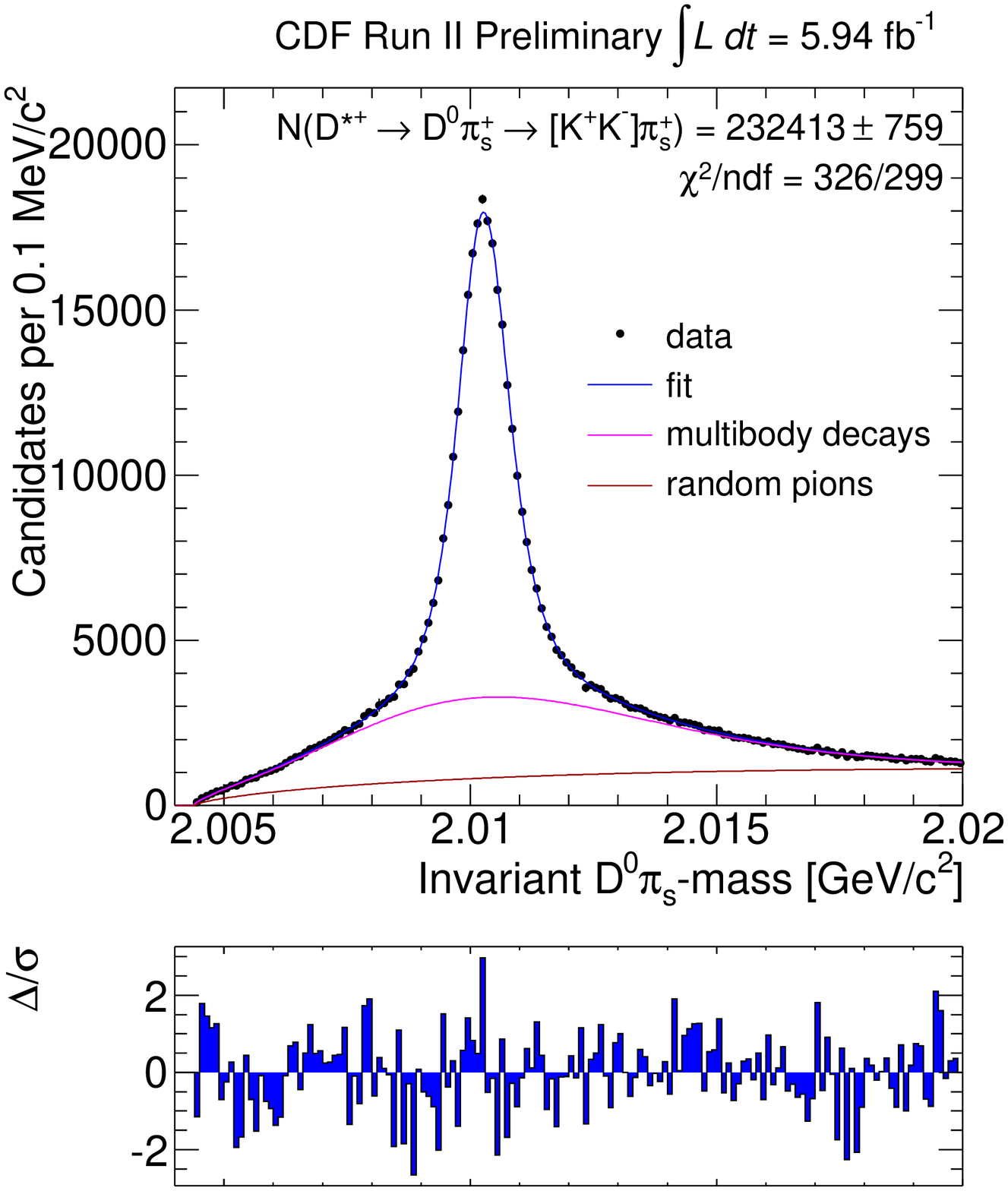}
\put(82,22){(c)}
\end{overpic}\hfil
\begin{overpic}[height=0.24\textheight,grid=false,trim=0 160pt 0 0,clip]{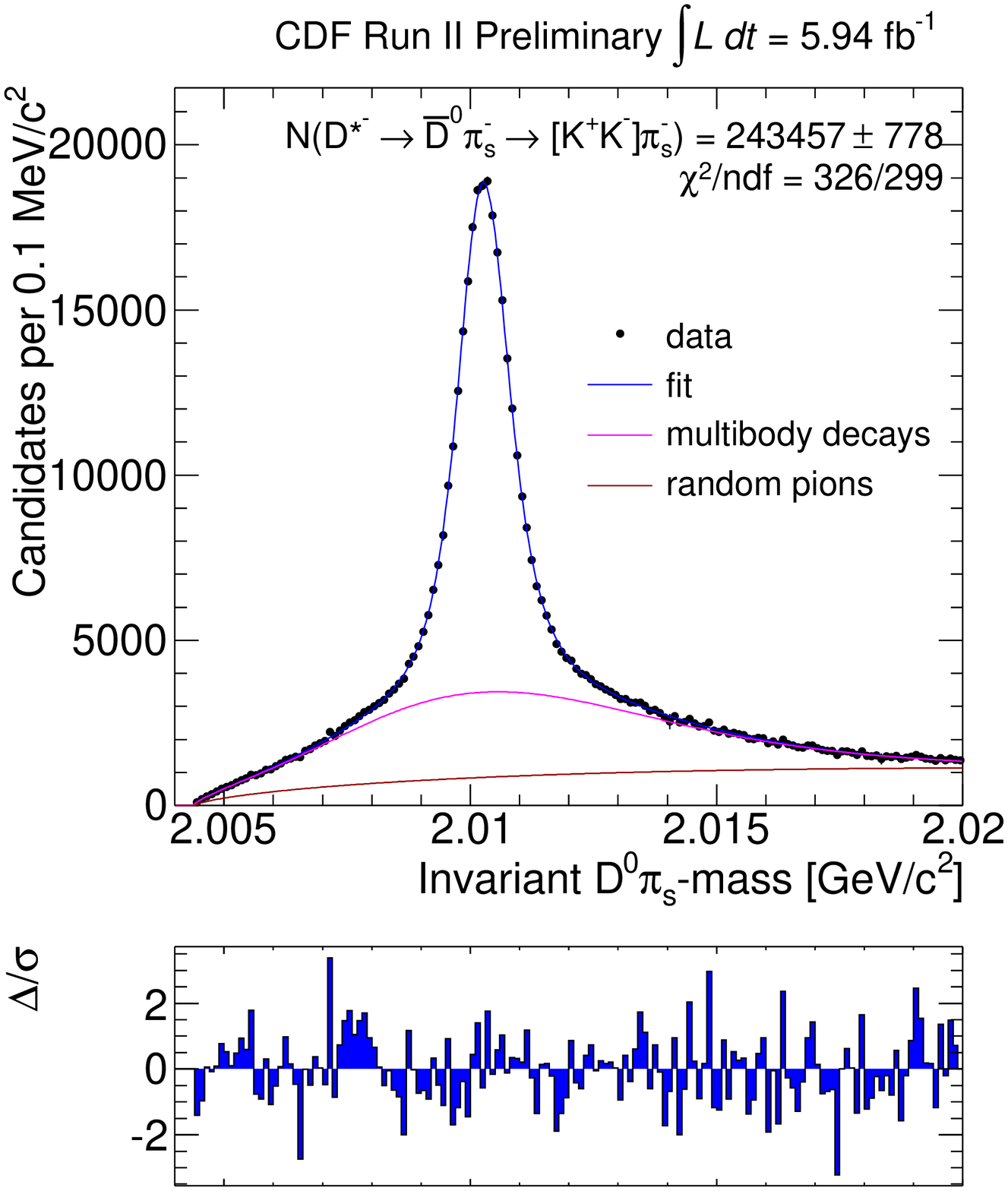}
\put(82,22){(d)}
\end{overpic}\vfill
\begin{overpic}[height=0.24\textheight,grid=false,trim=0 160pt 0 0,clip]{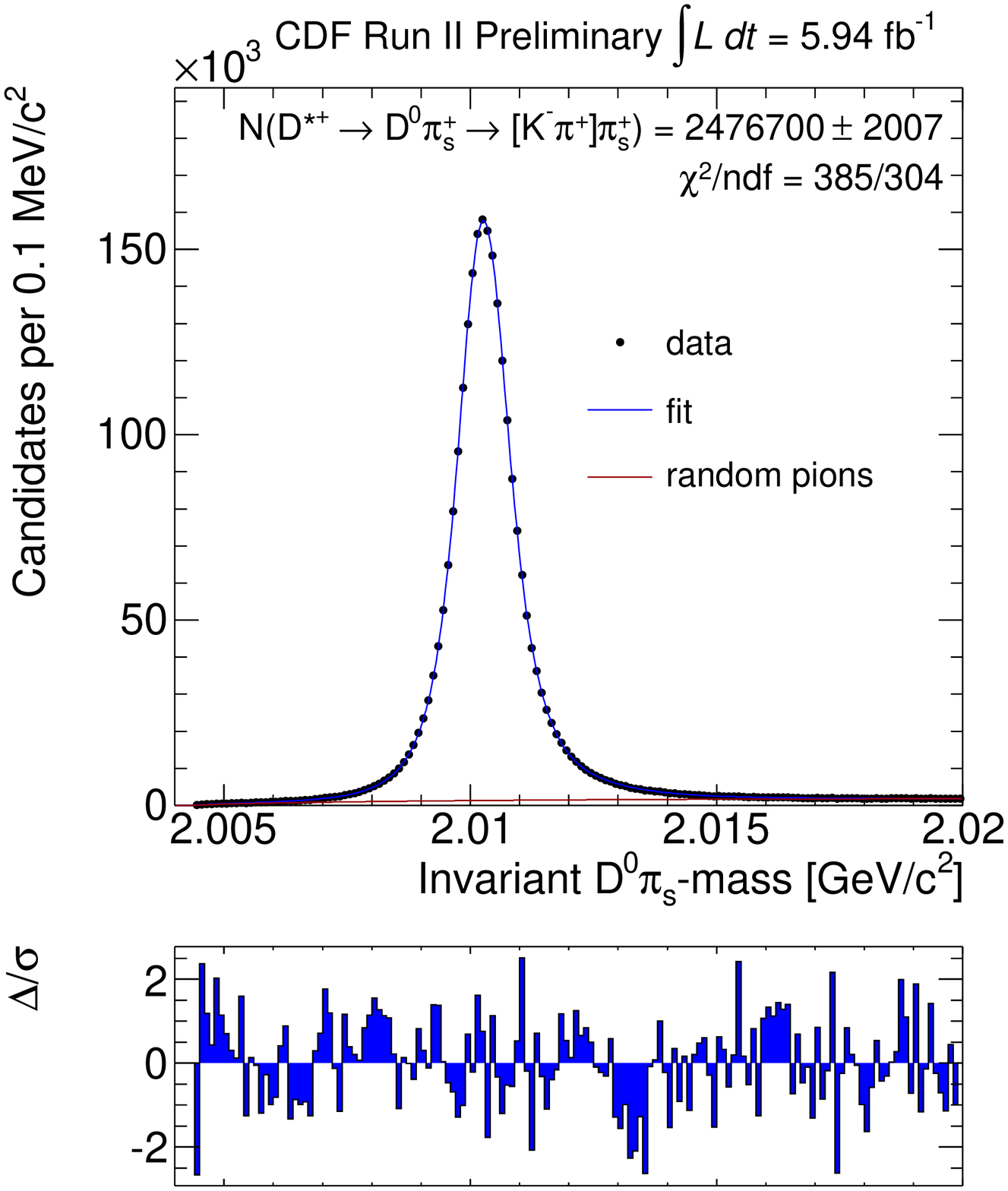}
\put(82,22){(e)}
\end{overpic}\hfil
\begin{overpic}[height=0.24\textheight,grid=false,trim=0 160pt 0 0,clip]{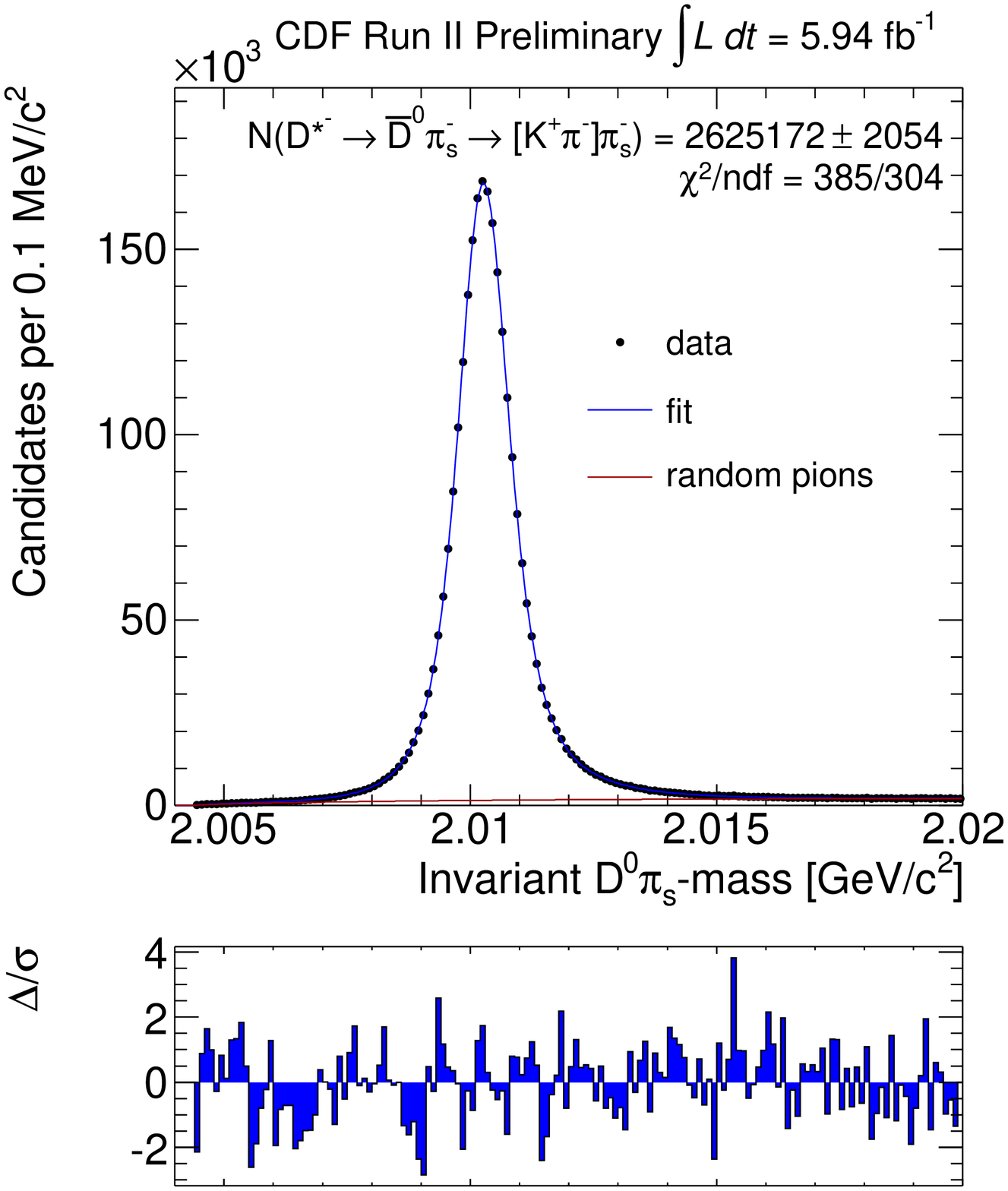}
\put(82,22){(f)}
\end{overpic}\vfill
\begin{overpic}[height=0.24\textheight,grid=false,trim=0 160pt 0 0,clip]{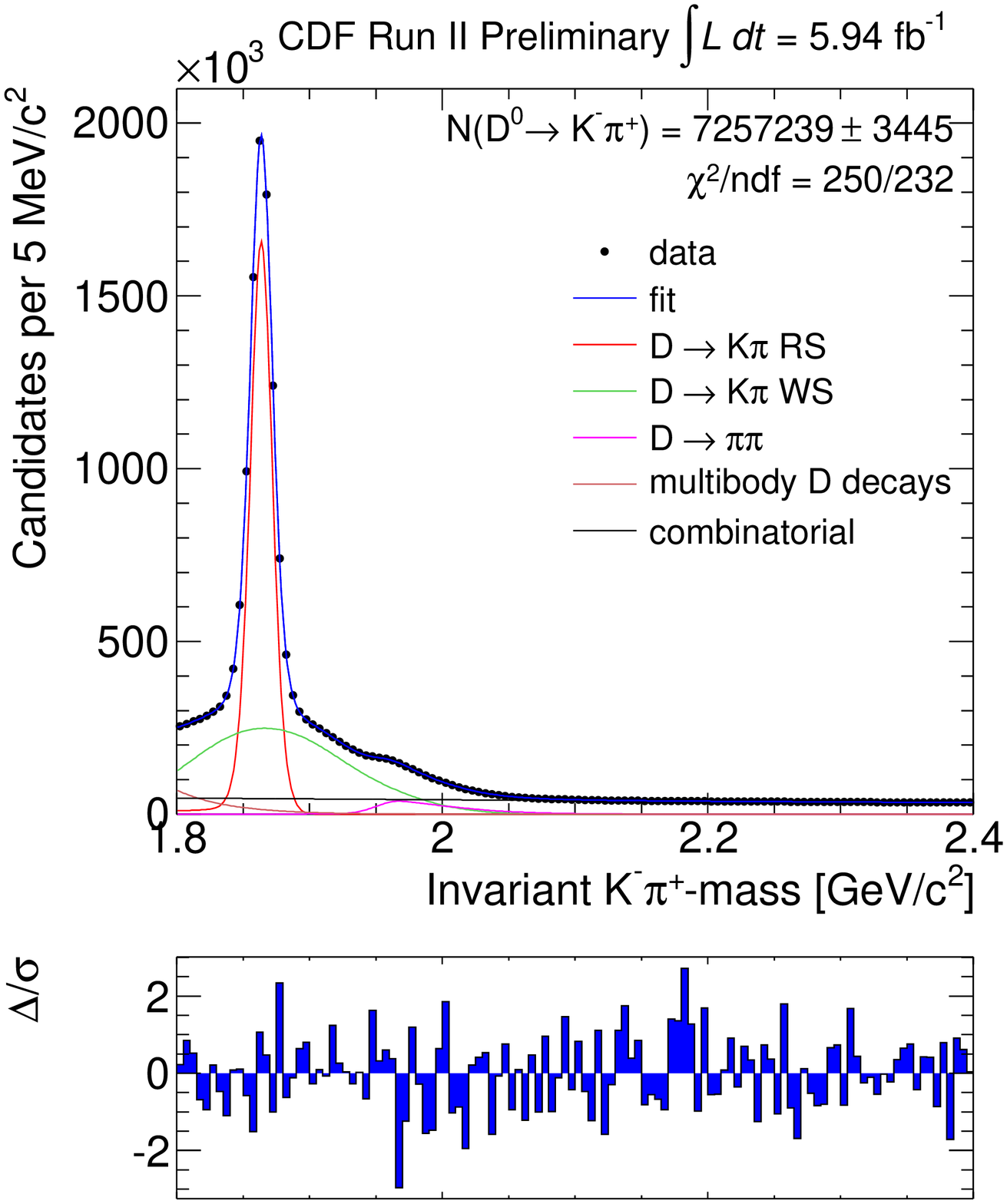}
\put(82,22){(g)}
\end{overpic}\hfil
\begin{overpic}[height=0.24\textheight,grid=false,trim=0 160pt 0 0,clip]{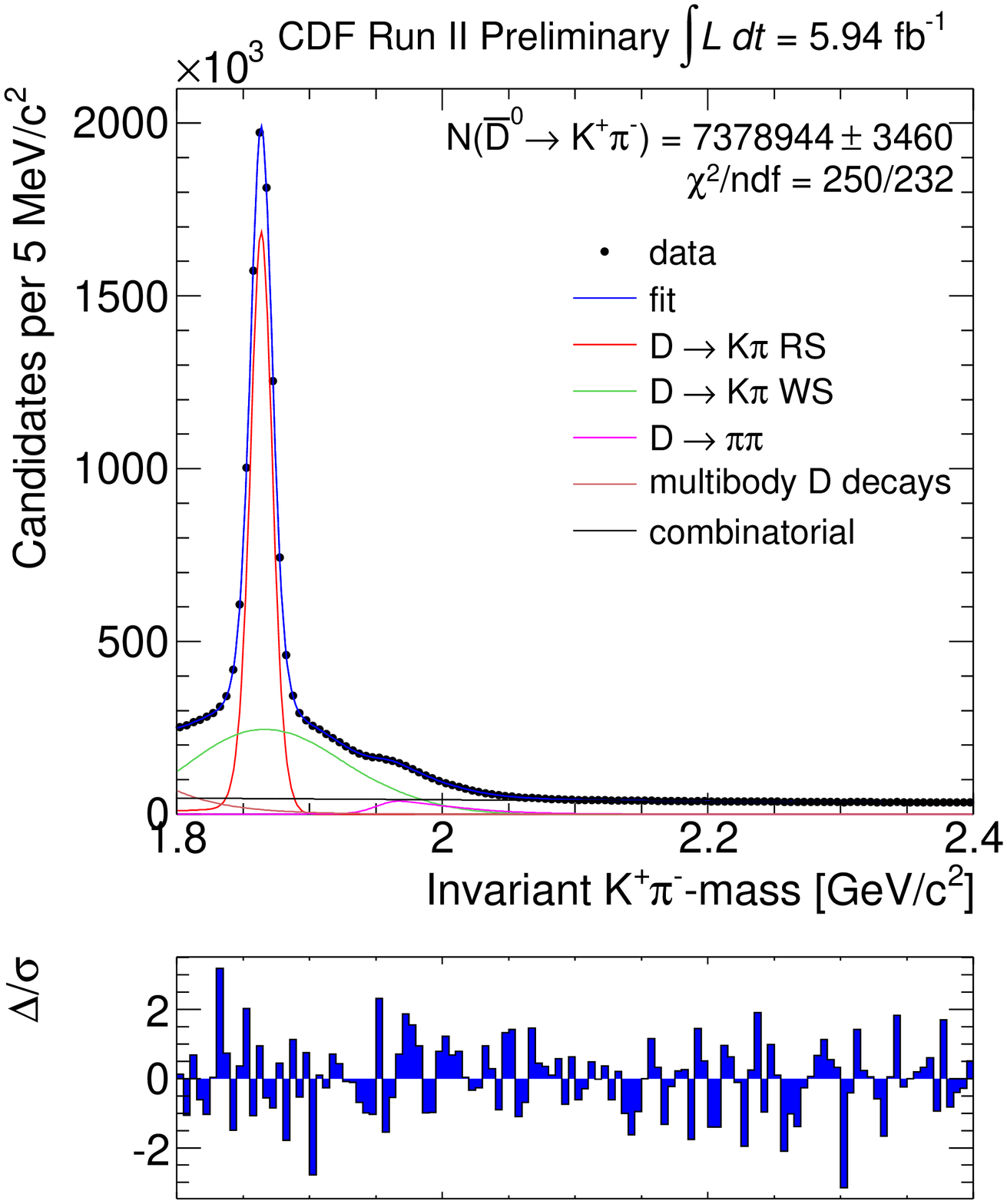}
\put(82,22){(h)}
\end{overpic}
\caption{Projections of the combined fit on data for tagged $D^0\to\pi^+\pi^-$ (a)-(b), tagged $D^0\to K^-K^+$ (c)-(d), tagged $D^0\to K^-\pi^+$ (e)-(f) and untagged $D^0\to K^-\pi^+$ (g)-(h) decays. Charm decays on the left and anticharm on the right.}\label{fig:fits}
\end{figure}

We extract independent signal yields for $D^0$ and $\Dbar^0$ candidates without using particle identification. In the $D^\star$-tagged samples we use the charge of the soft pion. In the untagged $D^0\to K^-\pi^+$ sample we randomly divided the sample in two independent subsamples similar in size. In each subsample we calculate the mass of each candidate with a specific mass assignment: $K^-\pi^+$ in the first subsample and $K^+\pi^-$ in the second one. In one sample the $D^0\to K^-\pi^+$ signal is correctly reconstructed and appears as a narrow peak (about $8$~MeV/c$^2$ wide), overlapping a broader peak of the misreconstructed $\Dbar^0\to K^+\pi^-$ component (red and green curves in figs.~\ref{fig:fits}~(g)-(h)). The reverse applies to the other sample. The yield asymmetry is extracted by fitting the number of candidates populating the two narrow peaks.

We determine the yields by performing a binned $\chi^2$ fit to the $D^0\pi_s$-mass ($K\pi$-mass) distribution combining charm and anticharm decays of both tagged (untagged) samples. The fits projections are shown in fig.~\ref{fig:fits}, the resulting raw asymmetries are: \mbox{$\Acp^{\text{raw}}(\pi\pi^\star)= (-1.86\pm0.23)\%$}, \mbox{$\Acp^{\text{raw}}(KK^\star)= (-2.32\pm0.21)\%$},\mbox{$\Acp^{\text{raw}}(K\pi^\star)=(-2.91\pm0.05)\%$}, \mbox{$\Acp^{\text{raw}}(K\pi)= (-0.83\pm0.03)\%$}.

\begin{table}[t]
\centering
\begin{tabular}{lcc}
\hline
Source of systematic uncertainty & $\Delta\Acp(\pi^+\pi^-)$ & $\Delta\Acp(K^+K^-)$ \\
\hline
Approximations in the method & $0.009\%$ & $0.009\%$\\
Beam drag effects & $0.004\%$ & $0.004\%$ \\
Contamination of non-prompt $D^0$ decays & $0.034\%$ & $0.034\%$ \\
Shapes used in fits & $0.010\%$ & $0.058\%$ \\
Shapes charge differences & $0.098\%$ & $0.052\%$ \\
Asymmetries from non-subtracted backgrounds & $0.018\%$ & $0.045\%$\\
\hline
Sum in quadrature & $0.105\%$ & $0.097\%$\\
\hline
\end{tabular}
\caption{Summary of systematic uncertainties.}\label{tab:syst}
\end{table}

The analysis has been tested using Monte Carlo samples simulated with a wide range of physical and detector asymmetries to verify that the cancellation, achieved by mean of eq.~\eqref{eq:formula}, works regardless of the specific configuration. These studies confirm the validity of our approach and provide a quantitative estimate of the systematic uncertainty coming from the basic assumptions in the method. All other systematic uncertainties are evaluated from data. A summary of all contributions to the final systematic error is shown in tab.~\ref{tab:syst}. Assuming they are independent and summing in quadrature we obtain a total systematic uncertainty on our final $\Acp(\pi\pi)$ ($\Acp(KK)$) measurement of $0.11\%$ ($0.10\%$), approximately half of the statistical uncertainty.

\section{Final results and conclusions}
We report a preliminary measurement of the CP asymmetry in the $D^0\to\pi^+\pi^-$ and $D^0\to K^+K^-$ decays using 5.94 fb$^{-1}$ of data collected by the CDF displaced track trigger. The final results are
\begin{align*}
\Acp(D^0\to\pi^+\pi^-) &= \bigl[+0.22\pm0.24\stat\pm0.11\syst\bigr]\%\qquad\text{and}\\
\Acp(D^0\to K^+K^-) &= \bigl[-0.24\pm0.22\stat\pm0.10\syst\bigr]\%,
\end{align*}
which are consistent with CP conservation and also with the SM predictions.

\begin{figure}[t]
\centering
\begin{overpic}[width=0.5\textwidth,grid=false]{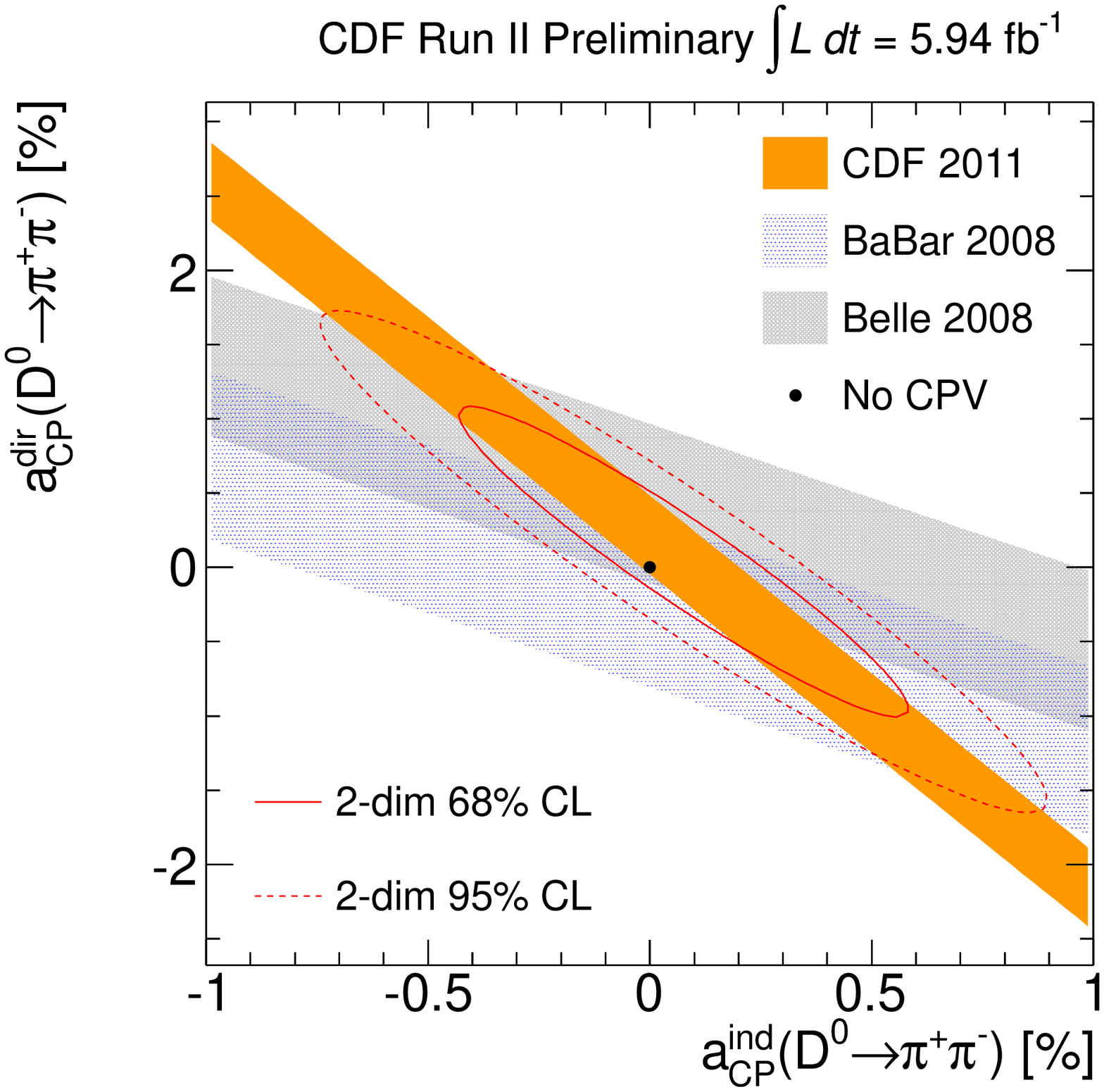}
\put(80,17){(a)} 
\end{overpic}\hfil
\begin{overpic}[width=0.5\textwidth,grid=false]{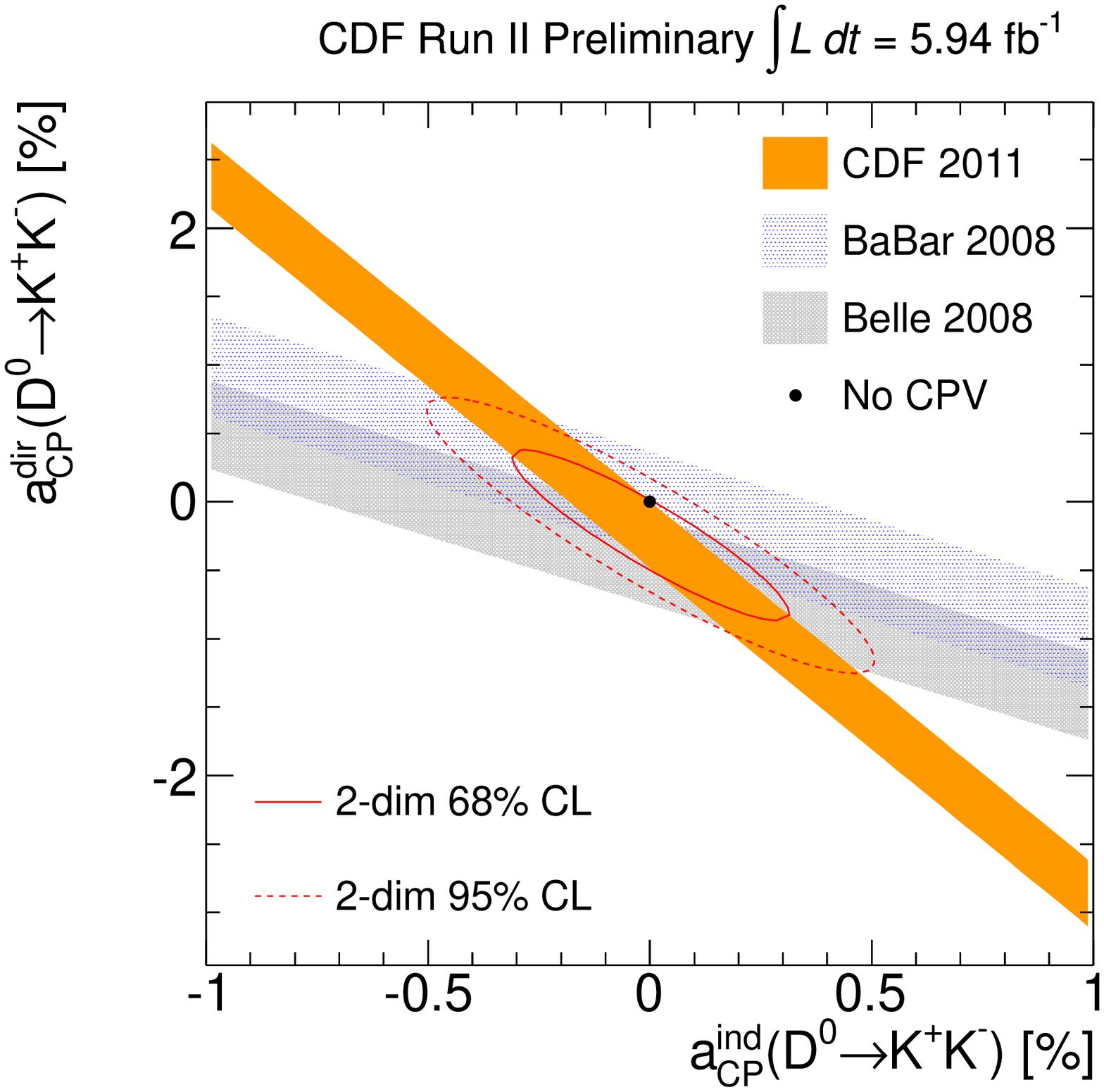}
\put(80,17){(b)}
\end{overpic}
\caption{Comparison of our measurements of the CP asymmetry in the $D^0\to \pi^+\pi^-$ (a) and $D^0\to K^+K^-$ (b) decays with current best results from B-factories in the parameter space $(\acp{ind},\acp{dir})$.}\label{fig:combination}
\end{figure}

As expressed by eq.~\eqref{eq:acp} the $\Acp(h^+h^-)$ measurement describes a straight line in the plane $(\acp{ind},\acp{dir})$ with angular coefficient given by $\langle t\rangle/\tau$. Because of a threshold on the impact parameter of tracks, imposed at trigger level, our sample of \mbox{$D^0\to\pi^+\pi^-$} (\mbox{$D^0\to K^+K^-$}) decays is enriched in higher-valued proper decay time candidates with a mean value of $2.40(2.65)\pm0.03\ (\mathit{stat.}+\mathit{syst.})$ times the $D^0$ lifetime, as measured from a fit to the proper time distribution. Due to their unbiased acceptance in charm decay time, B-factories samples have instead $\langle t\rangle\approx\tau$ \cite{babelle}. Hence, the combination of the three measurements allow to constrain independently both $\acp{dir}$ and $\acp{ind}$. Fig.~\ref{fig:combination} shows such combination: the bands are $1\sigma$ wide and the red curves represent the $68\%$ and $95\%$ CL regions of the combined result assuming Gaussian uncertainties.

\begin{figure}[t]
\centering
\begin{overpic}[width=0.4\textwidth,grid=false]{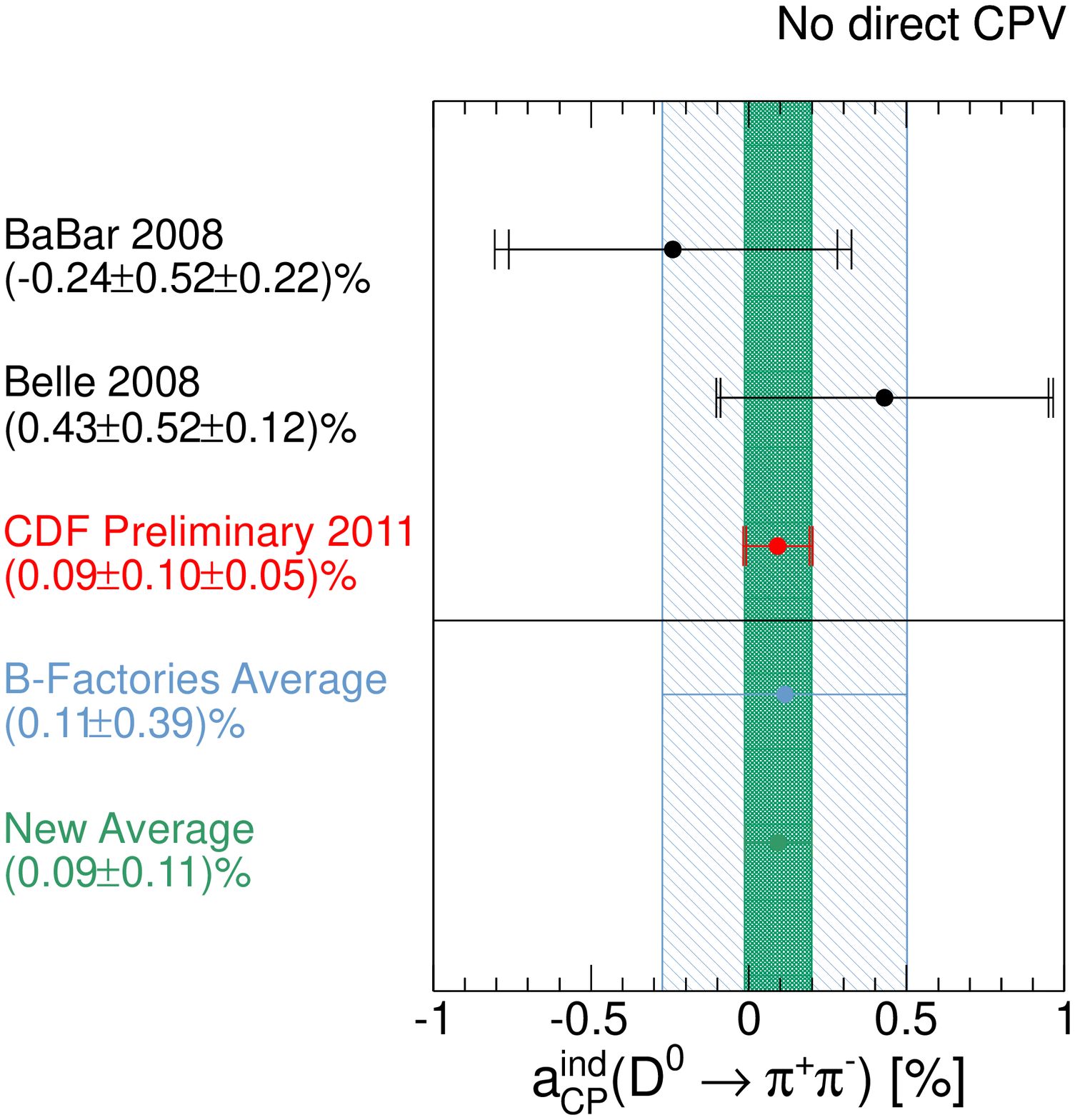}
\put(86,17){(a)} 
\end{overpic}\hfil
\begin{overpic}[width=0.4\textwidth,grid=false]{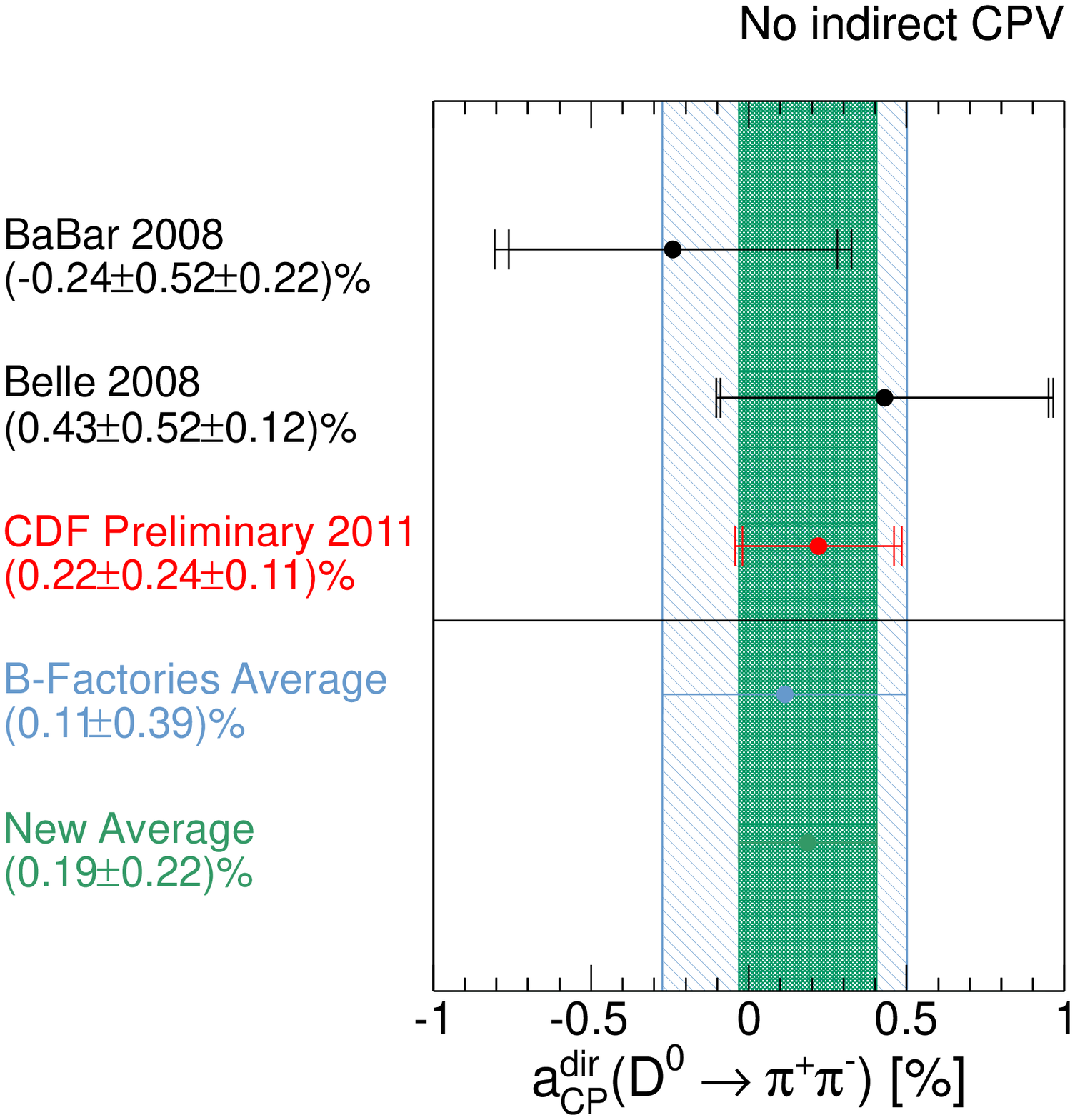}
\put(86,17){(b)}
\end{overpic}\\
\begin{overpic}[width=0.4\textwidth,grid=false]{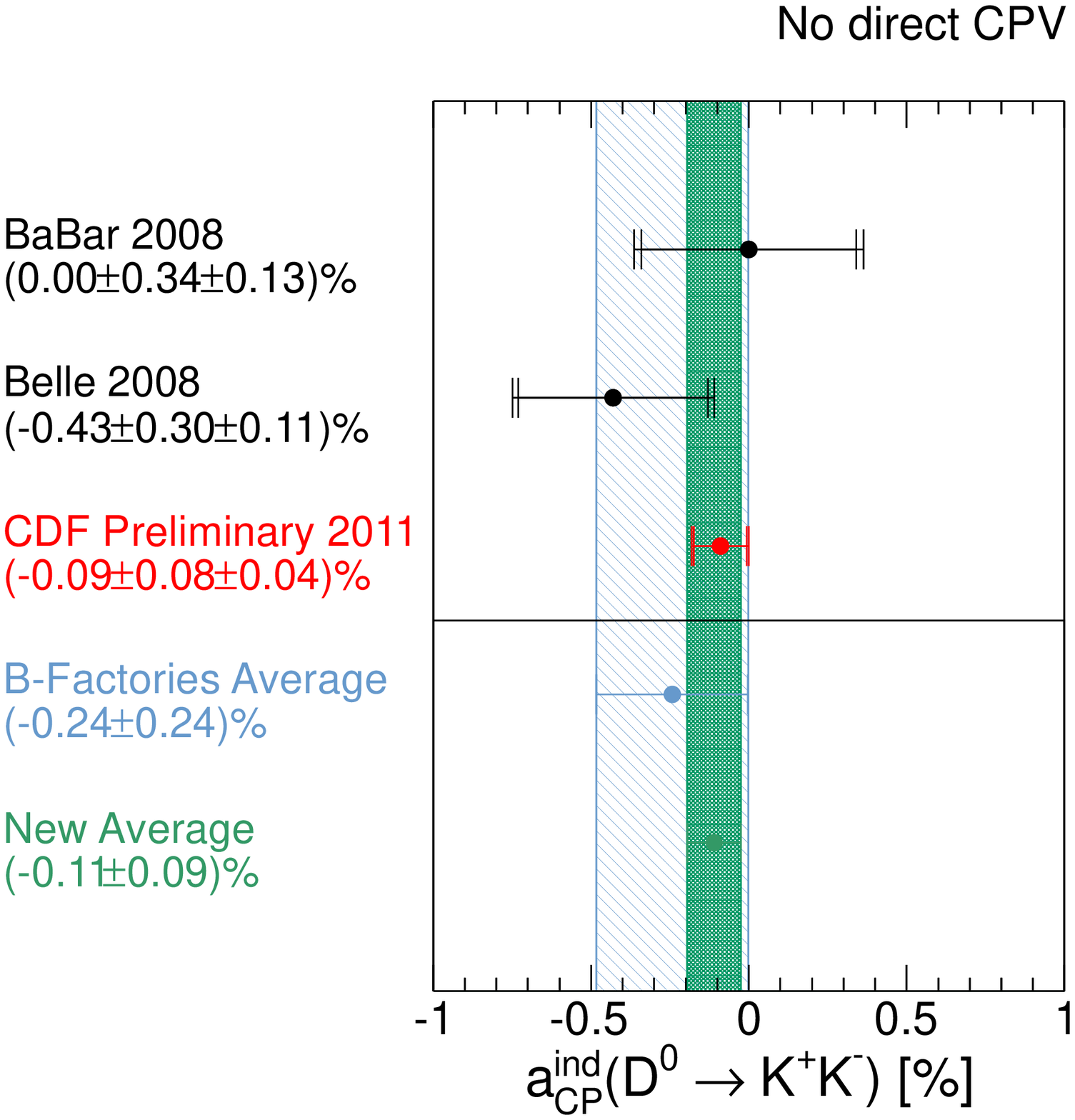}
\put(86,17){(c)}
\end{overpic}\hfil
\begin{overpic}[width=0.4\textwidth,grid=false]{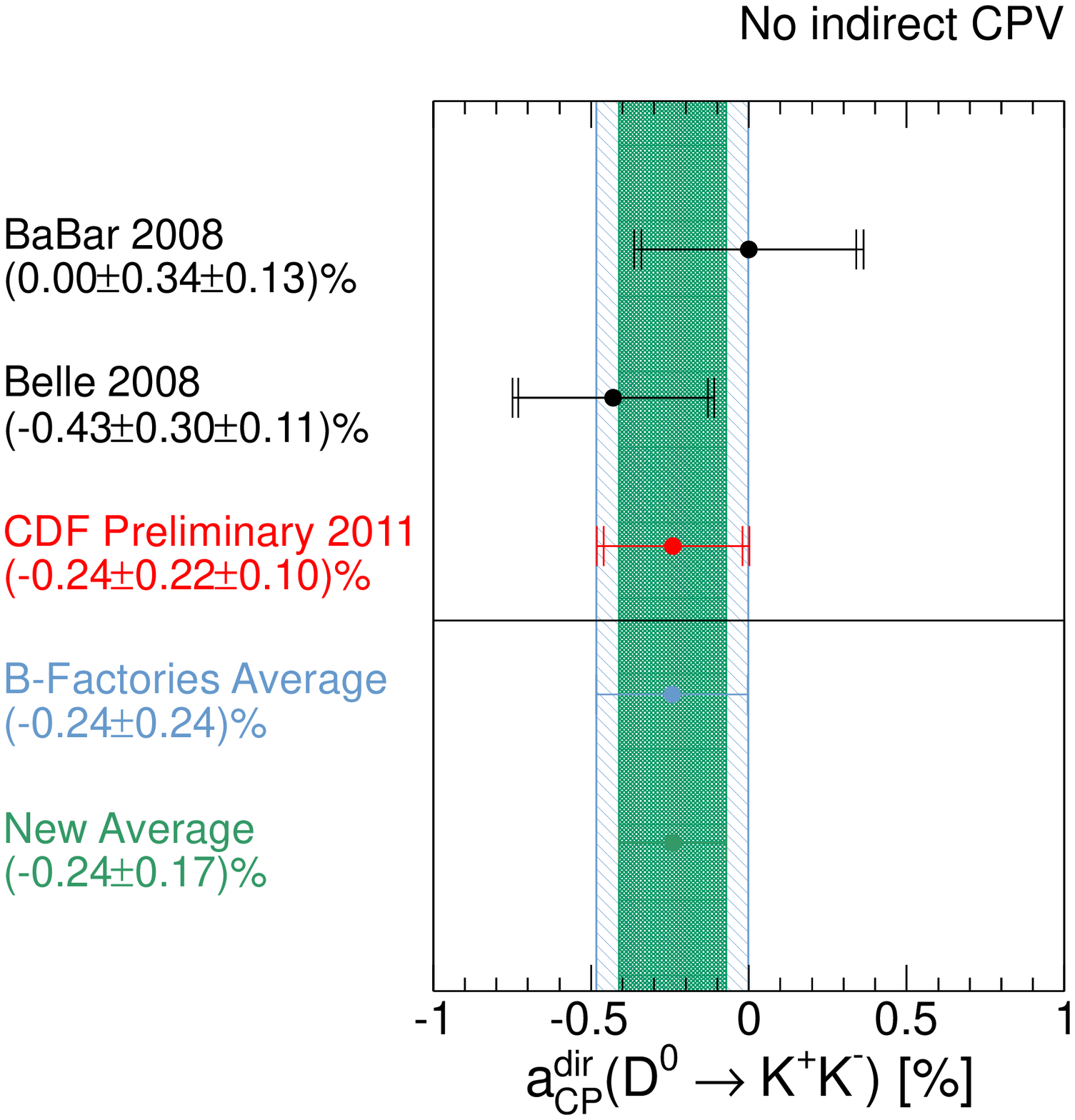}
\put(86,17){(d)}
\end{overpic}
\caption{Comparison of our measurements with B-factories experiments assuming no direct (a)-(c) or indirect (b)-(d) CP violation. In each plot the $1\sigma$ band of the average between B-factories measurements is represented in blue, while in green we report the new average computed including also these preliminary results.}\label{fig:direct_and_indirect}
\end{figure}

Assuming negligible direct CP violation in both decay modes, the observed asymmetry is only due to mixing, $\Acp(h^+h^-) \approx \acp{ind}\ \langle t \rangle / \tau$,  yielding
\begin{align*}
\acp{ind}(D^0\to\pi^+\pi^-) &= \bigl[+0.09\pm0.10\stat\pm0.05\syst\bigr]\%\quad\text{and}\\
\acp{ind}(D^0\to K^+K^-) &= \bigl[-0.09\pm0.08\stat\pm0.04\syst\bigr]\%.
\end{align*}
Because $\acp{ind}$ in this case is independent of the final state, the two measurements can be averaged,  assuming correlated systematic uncertainties,  to obtain a precise determination of CP violation in charm mixing:
$$\acp{ind}(D^0) = \bigl[-0.01\pm0.06\stat\pm0.05\syst\bigr]\%.$$
Fig.~\ref{fig:direct_and_indirect} shows the comparison with B-factories measurements in this hypothesis and conversely, assuming $\acp{ind}=0$.

In conclusion, we have measured the CP asymmetry in singly-Cabibbo suppressed $D^0$ decays with unprecedented precision and found no evidence for CP violation. These results probe significant regions of the space of parameters of charm phenomenology and are expected to provide a powerful discrimination between SM and different NP scenarios \cite{bigi}.

\end{document}